\documentstyle[12pt]{article}

\begin{document}

{\it University of Shizuoka}

\hspace*{9.5cm} {\bf US-94-02}\\[-.3in]

\hspace*{9.5cm} {\bf Revised }\\[-.3in]

\hspace*{9.5cm} {\bf June 1994}\\[.4in]

\begin{center}

{\large\bf Seesaw-Type Quark and Lepton Mass Matrices }\\[.1in]
{\large\bf and U(3)-Family Nonet Higgs Bosons}\\[.5in]

{\bf Yoshio Koide}\footnote{
E-mail: koide@u-shizuoka-ken.ac.jp} \\

Department of Physics, University of Shizuoka \\
395 Yada, Shizuoka 422, Japan \\[.1in]

and \\[.1in]

{\bf Hideo Fusaoka}\footnote{
E-mail: fusaoka@amugw.aichi-med-u.ac.jp} \\

Department of Physics, Aichi Medical University \\
Nagakute, Aichi 480-11, Japan \\[.5in]

{\large\bf Abstract}\\[.1in]

\end{center}

\begin{quotation}
A unified mass matrix model of quarks and leptons with a seesaw-type form
$M_f=m M_F^{-1} m$ is proposed on the basis of a non-standard Higgs scenario.
The matrix $m$ is provided by U(3)-family nonet bosons $\phi$,
and the matrix $M_F$ is a mass matrix of heavy fermions $F_i$
corresponding to the ordinary fermions $f_i=\nu_i,e_i,u_i,d_i$ ($i=1,2,3$).
It is shown that a Higgs potential of $\phi$ with a broken U(3)-family
symmetry leads to a desirable charged lepton mass formula when
$M_E\propto {\bf 1}$.
Then, phenomenologically desirable forms of heavy quark mass matrices
$M_Q$ ($Q=U,D$) are investigated.
\end{quotation}

\newpage

Recent measurements  of tau lepton mass $m_\tau$ have found [1] that
the observed value  excellently satisfies a charged lepton mass formula
$$
m_e+m_\mu+m_\tau=\frac{2}{3}(\sqrt{m_e}+\sqrt{m_\mu}+\sqrt{m_\tau})^2
\ , \eqno(1)
$$
which provides  $m_\tau=1777$ MeV for the input values of $m_e$ and $m_\mu$.
The relation (1) has first been speculated on the basis of
a composite model [2] and then an extended
technicolor-like model [3] by one of the authors (Y.K.).
In those models, it is essential to assume that the mass matrix $M_f$ for
fermions $f_i$ ($i=1,2,3$: family indices) is given by the form
$$ M_f = m_0^f \, G O_f G \ ,  \eqno(2) $$
where $ G = {\rm diag}(g_1, g_2, g_3)$.
If we assume that the parameters $g_i$  satisfy the relations
$g_i=g_i^{(8)}+g^{(1)}$, $\sum_i g_i^{(8)}=0$ and
$\sum_i\left(g_i^{(8)}\right)^2=3\left(g^{(1)}\right)^2$,
and the matrix form of $O_f$ is given by $O_e={\bf 1}$
({\bf 1} is a $3\times 3$ unit matrix)
for the charged lepton mass matrix $M_e$,
then we can obtain the relation (1).
Furthermore, one of the authors (Y.K.) has recently pointed out [4] that
the mass matrix (2) has a possibility that it can provide
a unified description of quark and lepton mass matrices by considering
a special form of $O_f$ (without re-fitting the parameters $g_i$
for each $f=u,d,\nu$).

In Ref.[3], the matrix form (2) was speculated on the basis of
an extended technicolor-like model.
Then, in that model, there were too many fermions,
because we must consider many technicolored fermions $F_{i\alpha}$
($\alpha$ is technicolor index)
corresponding to the ordinary fermions $f_i$.
Therefore, in that model, SU(3)-color cannot be asymptotic free.

In the present paper, we propose an alternative model based on
a non-standard Higgs model,
which leads to a seesaw-type mass matrix $M_f\simeq m M_F^{-1}m$ and
in which U(3)-family nonet Higgs scalars $\phi$ play an essential role of
deriving the mass formula (1).
First, in order to give an outline of the model,
we will discuss the case for charged leptons, where $M_F$ ($F=E$)
is simply assumed as $M_E\propto {\bf 1}$.
Next, we will discuss possible forms of the heavy fermion mass matrices
$M_F$ ($F=U,D,N$) from the phenomenological point of view (but within
the context of our Higgs scenario).

In our scenario, we prepare the following fermions:
$f=\ell, q$ ($\ell=(\nu,e)$, $q=(u,d)$) and
$F=N, E, U, D$, which belong to $f_L=({\bf 2},{\bf 1},{\bf 3})$,
$f_R=({\bf 1},{\bf 2},{\bf 3})$, $F_L=({\bf 1},{\bf 1},{\bf 3})$, and
$F_R=({\bf 1},{\bf 1},{\bf 3})$ of
SU(2)$_L\times$SU(2)$_R\times$U(3)$_{family}$, respectively.
Up- and down-heavy fermions, $F^{up}$ and $F^{down}$, are distinguished
by hypercharge  $Y$  (note that $Y\neq B-L$ for the heavy fermions):
Hypercharges of the heavy fermions $(N,E)$ and $(U,D)$ take the values
$(0,-2)$ and $(4/3,-2/3)$, respectively.
In the present model, differently from the standard
SU(2)$_L\times$SU(2)$_R\times$U(1)$_{B-L}$ model,
we do not consider Higgs scalar fields which
belong to $({\bf 2}, {\bf 2})$ of SU(2)$_L\times$SU(2)$_R$,
so that there are no Higgs fields which couple with
$\overline{f}f$ at tree level.
We assume only the following Yukawa interactions:
$$
H_{Yukawa}=g_F \sum_{i,j}\overline{F}^i(\Phi_F)_i^j F_j
+g_L \sum_{i,j}\left(\overline{f}_L^i(\phi_L)_i^j F_{Rj}^{down}
+\overline{f}_L^i(\widetilde{\phi}_L)_i^j F_{Rj}^{up} + H.C.\right)
+(L\leftrightarrow R) \ , \eqno(3)
$$
where $\phi=(\phi^+,\phi^0)$ and
$\widetilde{\phi}=(\overline{\phi}^0,-\phi^-)$.
The scalar fields $\phi_L$ and $\phi_R$ belong to
$({\bf 2}, {\bf 1},{\bf 8}+{\bf 1})$ and $({\bf 1}, {\bf 2},{\bf 8}+{\bf 1})$
of SU(2)$_L\times$SU(2)$_R\times$U(3)$_{family}$, respectively,
and the vacuum expectation values (VEV's) $\langle \phi^0_L\rangle$ and
$\langle \phi^0_R\rangle$ provide left- and right-handed weak boson masses
$m(W_L)$ and $m(W_R)$, respectively.
The fields $\Phi_F$ which belong to ({\bf 1},{\bf 1},{\bf 8}+{\bf 1}) do not
contribute to weak boson masses $m(W_L)$ and $m(W_R)$, but play only
a role of providing extremely large masses for vector-like fermions $F$.
(The structure of $\langle\Phi_F\rangle$ will be discussed later.)
Under the approximation of $M_F\gg m_L, m_R$ ($M_F=g_F \langle\Phi_F\rangle$,
$m_L=g_L \langle \phi^0_L\rangle$, and so on), we obtain a seesaw-type
mass matrix form $M_f\simeq m_L M_F^{-1}m_R$.


First, we discuss the charged lepton mass matrix $M_e$.
We assume that charged heavy leptons $E$ couple only with
a U(3)$_{family}$ singlet scalar field  $\Phi_0$,
which belongs to $({\bf 1}, {\bf 1},{\bf 1})$
of SU(2)$_L\times$SU(2)$_R\times$ U(3)$_{family}$,
so that the matrix form $O_e$ in (2) is given by $O_e={\bf 1}$.
On the other hand, for the scalar fields $\phi_L$, we assume
the following Higgs potential which is approximately invariant
under the U(3)-family symmetry:
$$
V(\phi)=\mu^2 {\rm Tr}\left( \phi^-\phi^+
+\overline{\phi}^0\phi^0 \right)
+\frac{1}{2}\lambda \left[{\rm Tr}\left( \phi^-\phi^+
+\overline{\phi}^0 \phi^0\right) \right]^2 $$
$$+\frac{1}{2}\lambda' \left[{\rm Tr}\left( \phi^-\phi^-\right)
{\rm Tr}\left( \phi^+\phi^+\right)
+2{\rm Tr}\left( \phi^-\overline{\phi}^0\right)
{\rm Tr}\left( \phi^+\phi^0\right)
+{\rm Tr}\left(\overline{\phi}^0\overline{\phi}^0\right)
{\rm Tr}\left( \phi^0\phi^0\right)\right] $$
$$+\eta \left( \phi^-_1\phi^+_1
+\overline{\phi}^0_1\phi^0_1\right)
{\rm Tr}\left( \phi^-_8\phi^+_8 +\overline{\phi}^0_8\phi^0_8 \right)
+\eta' {\rm Tr}\left[\left( \phi^-_1\phi^+_8
+\overline{\phi}^0_1\phi^0_8 \right)
\left( \phi^-_8\phi^+_1 +\overline{\phi}^0_8\phi^0_1 \right)\right] \ ,
\eqno(4)
$$
where, for convenience,  we have omitted the index $L$.
Here, the traces are taken over the family indices and
$\phi_8=(\phi^+_8, \phi^0_8)$ and $\phi_1=(\phi^+_1, \phi^0_1)$
denote octet and singlet components of $\phi$, respectively:
$\phi=\phi_8 +(1/\sqrt{3})\phi_1 {\bf 1}$.
In the potential (4), the U(3)-family invariance of $V(\phi)$
is explicitly broken by  the $\eta$- and $\eta'$-terms,
while an SU(3)-family symmetry is still unbroken.
Hereafter, we sometime use the language of SU(3)-family instead of
broken U(3)-family.
Of course, the potential (4) is not a general form of the
SU(2)$_L\times$SU(2)$_R\times$SU(3)$_{family}$ invariant potential.

For $\mu^2<0$, conditions for minimizing the potential $V(\phi)$
are as follows:
$$
\left[\mu^2-\lambda {\rm Tr}({\bf v}^\dagger {\bf v})
-(\eta+\eta'){\rm Tr}({\bf v}_8^\dagger {\bf v}_8)\right] v_1^*
-\lambda'{\rm Tr}({\bf v}^\dagger {\bf v}^\dagger) v_1=0 \ , \eqno(5a)
$$
$$
\left[\mu^2-\lambda {\rm Tr}({\bf v}^\dagger {\bf v})
-(\eta+\eta') v_1^* v_1 \right] ({\bf v}_8^\dagger)_i^j
-\lambda'{\rm Tr}({\bf v}^\dagger {\bf v}^\dagger)
({\bf v}_8)_i^j = 0 \ , \eqno(5b)
$$
and equations exchanged as (${\bf v}\leftrightarrow {\bf v}^\dagger$,
${\bf v}_8\leftrightarrow {\bf v}_8^\dagger$, and $v_1\leftrightarrow v_1^*$)
in (5$a$) and (5$b$), respectively,
where ${\bf v}=\langle \phi^0\rangle$, ${\bf v}_8=\langle \phi^0_8\rangle$
and $v_1=\langle \phi^0_1\rangle=({\rm Tr}\, {\bf v})/\sqrt{3}$.
These conditions lead to the relations
${\bf v}^\dagger={\bf v}$ and
$$v_1^*v_1 = {\rm Tr}\left( {\bf v}_8^\dagger {\bf v}_8 \right)
={-\mu^2}/\left[2(\lambda +\lambda')+\eta+\eta'\right] \ ,$$
so that we obtain the relation
$$
{\rm Tr}\left( {\bf v}^2\right) =
\frac{2}{3} \left({\rm Tr}\, {\bf v}\right)^2
\ . \eqno(6)
$$

We assume that a Higgs potential $V(\phi_R)$ has the same structure
with $V(\phi_L)$, i.e., each term in $V(\phi_R)$ takes the coefficient
which is exactly proportional to the corresponding term in $V(\phi_L)$.
This assumption means that there is a kind of ``conspiracy" between
$V(\phi_R)$ and $V(\phi_L)$.
However, in this paper, we do not go into this problem moreover.
When we assume $\langle\phi_R^0\rangle\propto\langle\phi_L^0\rangle$,
we can obtain the mass formula (1) from the relation (6).

Re-derivation of the relation (1) on the basis of a Higgs potential
with a broken U(3)-family symmetry has also been done
by one of the authors (Y.K.) [5].
However, since his potential lacked the $\lambda'$- and $\eta'$-terms
in (4), too many massless Nambu-Goldstone states appeared,
although he successfully derived the relation (6).
In order to avoid such excess of massless states, we need
the $\lambda'$-term.
On other hand, if we add, for example, a term
Tr$\left[ (\phi^-\phi^++\overline{\phi}^0\phi^0)^2\right]$
to the potential (4), we cannot obtain the relation (6).
In order to the relation (6), the potential $V(\phi)$ is restricted
to a special form under the broken U(3)-family symmetry.

Next, we discuss the form of $O_f$.
In Ref.[4], it was pointed out that a form
$O_f={\bf 1} +3 a_f X(\phi_f)$ can provide successful predictions of
quark masses and Kobayashi-Maskawa (KM) [6] matrix elements,
where the matrix $X(\phi_f)$ is a democratic type matrix with
a phase factor $\phi_f$,
$$
X(\phi)=\frac{1}{3} \, \left(
\begin{array}{ccc}
1 & e^{i\phi} & 1 \\
e^{-i\phi} & 1 & 1 \\
1 & 1 & 1
\end{array} \right) \ . \eqno(7)
$$
However, in the context of a Higgs scenario,
it is not so easy to derive the matrix form of the second term
$X(\phi_f)$, which is not a rank one matrix for $\phi_f\neq 0$.
In the present paper, we consider alternative form of $O_f$.

We assume that the U(3)$_{family}$ singlet field $\Phi_0$ couples
with all heavy fermions $F$ universally.
In addition to $\Phi_0$, we assume a Higgs field $\Phi_X$ which
couples only with a heavy
quark states $F_S=(F_1+F_2+F_3)/\sqrt{3}$ ($F_i=U_i,D_i$)
(a symmetric representation of the permutation group S$_3$ [7]).
The VEV  $\langle\Phi_X\rangle$ provides a democratic mass matrix term
without phase factors, $X(0)$, which is a rank one matrix.
We consider that the coupling constants of $\Phi_X$ with heavy fermions
$F$ are different according as $F=U$ or $F=D$.
Therefore, the matrix form $O_f$ in (2) is given by
$$
O_f={\bf 1}+3a_f e^{i\alpha_f} X(0) \ , \eqno(8)
$$
i.e.,
$$
M_F\propto O_f^{-1}={\bf 1}+3b_f e^{i\beta_f} X(0) \ , \eqno(9)
$$
where
$$
b_f e^{i\beta_f} =- a_f e^{i\alpha_f}/(1+3a_f e^{i\alpha_f}) \ . \eqno(10)
$$
The reason that we still consider a democratic matrix from in $O_f$ is
motivated by only a phenomenological reason suggested in Ref.[4],
i.e., by the fact that for up-quark mass matrix with $\beta_u=0$,
we can obtain the successful mass relation [4]
$${m_u}/{m_c} \simeq {3m_e}/{4m_\mu} \ , \eqno(11)$$
for a small value of $\varepsilon_u\equiv 1/a_u$.
Note that the ratio $m_u/m_c$ is insensitive to the parameter $a_u$.
The parameter $\varepsilon_u\equiv 1/a_u$ is determined by the mass relation
$${m_c}/{m_t} \simeq 2(m_\mu/m_\tau)\varepsilon_u \ . $$

Differently from the model given in Ref.[4], down-quark mass matrix
$M_d$ with $\alpha_d\neq 0$ in the present model is not Hermitian.
We will demonstrate that the present model with the form (8) also
can provide reasonable predictions of quark mass ratios and KM matrix
by adjusting our parameters $a_u$, $a_d$ and $\alpha_d$.

In the present model, a case $a_d\simeq -0.5$ can provide
phenomenologically interesting predictions as seen below.
For small values of $|\alpha_d|$ and $\varepsilon_d\equiv -(2+a_d^{-1})$,
we obtain the down-quark mass ratios
$$
{m_s}/{m_b} \simeq ({1}/{2}) \kappa\left(1-
48\sqrt{2\varepsilon/3}\right) \ , \eqno(12)
$$
$$
{m_d}/{m_s} \simeq {16}(\varepsilon/{\kappa^2})
\left(1+96\sqrt{2\varepsilon/3}\right) \ , \eqno(13)
$$
where
$$
\kappa = \sqrt{\sin^2\frac{\alpha_d}{2} +
\left(\frac{\varepsilon_d}{4}\right)^2} \ , \ \
\varepsilon = \frac{m_e m_\mu}{m_\tau^2} \ . \eqno(14)
$$
We also obtain
$$
{m_d m_s}/{m_b^2}\simeq 4 {m_e m_\mu}/{m_\tau^2} \ , \eqno(15)
$$
as a relation which is insensitive to the small parameters $|\alpha_d|$
and $\varepsilon_d$.

Furthermore, we can obtain ratios of up-quarks to down-quarks,
for example,
$$
{m_u}/{m_d} \simeq 6 \kappa \sim 12 {m_s}/{m_b} \ . \eqno(16)
$$
Suitable choice of small values of $\varepsilon_d$ and
$\alpha_d$ ensures $m_u/m_d\sim O(1)$ in spite of $m_t\gg m_b$.
{}From (10), a small value $|\varepsilon_u|=1/|a_u|\simeq 0$ means
$b_u\simeq -1/3$, while a small value
$|\varepsilon_d|=|2+a_d^{-1}|\simeq 0$ means $b_d\simeq -1$.
It is noted that, in spite of the large ratio of $m_t/m_b$,
the ratio of $b_d/b_u$ is not so large, i.e., $b_d/b_u\simeq 3$.


Then, let us discuss the KM matrix elements $V_{ij}$.
The KM matrix $V$ is given by
$V= U_L^u P U_L^{d\dagger}$,
where $U_L^u$ and $U_L^d$ are defined by
$U_L^u M_u M_u^\dagger U_L^{u\dagger}={\rm diag}(m_u^2, m_c^2, m_t^2)$ and
$U_L^d M_d M_d^\dagger U_L^{d\dagger}={\rm diag}(m_d^2, m_s^2, m_b^2)$,
respectively, and $P$ is a phase matrix.
Here, we have considered that the quark basis for the mass matrix (2) can,
in general, deviate from the quark basis of weak interactions
by some phase rotations,
The simplest case $P=$diag$(1,1,1)$ cannot provide reasonable predictions
of $|V_{ij}|$.
Only when we take $P={\rm diag}(1,1,-1)$,
we can obtain reasonable predictions for
both quark mass ratios and KM matrix elements,
although it is an open question why such a phase inversion is caused on
the third family quark.
The predictions of $|V_{ij}|$ are sensitive to every values of
$\varepsilon_u$, $\varepsilon_d$ and $\alpha_d$, so that it is not
adequate to express $|V_{ij}|$ as simple approximate relations
such as those in (11)--(16).
Therefore, we will show only numerical results for $|V_{ij}|$.
For example, by taking $b_u=-0.3295$, $b_d=-1.072$, $\beta_u=0$ and
$\beta_d=18.5^\circ$, which are chosen by fitting the quark mass ratios,
we obtain the following predictions of quark masses, KM matrix elements
$|V_{ij}|$ and the rephasing-invariant quantity [8] $J$:
$$
\begin{array}{lll}
m_u=0.00228\ {\rm GeV} \ , & m_c=0.591 \ {\rm GeV} \ , &
m_t=170 \ {\rm GeV} \ , \\
m_d=0.00429\ {\rm GeV} \ , & m_s=0.0875 \ {\rm GeV} \ , &
m_b=3.02 \ {\rm GeV} \ ,  \\
\end{array}   \eqno(17)
$$
$$
|V_{us}|=0.223 \ , \ \ \ |V_{cb}|=0.0542 \ , \ \ \ |V_{ub}|=0.00309 \ ,
\ \ \ |V_{td}|=0.0146 \ , $$
$$ |V_{ub}/V_{cb}|=0.0570 \ , \ \ \ J=2.30 \times 10^{-5} \ . \eqno(18)
$$
The prediction $|V_{cb}|=0.0542$ in (18) is somewhat large in comparison
with the experimental value $|V_{cb}|=0.043\pm 0.007$ [9].
If we use $P=(1,1,-e^{i\delta})$ with a small phase value $\delta$
instead of $P=(1,1,-1)$,
we can obtain more excellent predictions without changing
predictions of quark masses in (17):
for example, when we take $\delta=-3.4^\circ$, we obtain
$|V_{us}|=0.223$, $|V_{cb}|=0.0431$, $|V_{ub}|=0.00282$
($|V_{ub}/V_{cb}|=0.0654$), $|V_{td}|=0.01184$ and $J=1.69\times 10^{-5}$.

In the numerical predictions of quark masses, (17),
we have used a common enhancement factor
of quark masses to lepton masses, $m_0^u/m_0^e=m_0^d/m_0^e=3$,
in order to compare with
quark mass values [10] at the energy scale
$\mu=\Lambda_W\equiv ({\rm Tr}\langle\phi_L^0\rangle^2)^{1/2}/\sqrt{2}=
(\sqrt{2}G_F)^{-1/2}/\sqrt{2}=174$ GeV:
$m_u=0.0024\pm 0.0005$ GeV, $m_d=0.0042\pm 0.0005$ GeV,
$m_c=0.605\pm 0.009$ GeV, $m_s=0.0851\pm 0.014$ GeV,
$m_t=174\pm 16$ GeV, and $m_b=2.87\pm 0.03$ GeV,
where we have used [9] $\Lambda_{\overline{MS}}^{(4)}=0.26$ GeV.
Although we are happy if we can explain such the factor $m_0^q/m_0^e=3$
by evolving quark and lepton masses from $\mu=\Lambda_X$
(a unification scale [11]) to $\mu=\Lambda_W$,
unfortunately, it is not likely to derive such a large factor $\sim 3$
from the conventional renormalization calculation.
We must assume an additional enhancement mechanism, for example,
a different coupling strength of $\Phi_0$ with heavy quarks
from that with heavy leptons.


For neutrinos, if we take $b_\nu\simeq -1/3$ and $\beta_\nu=0$
similar to the case of up-quarks,
we can obtain an interesting prediction [12]
for the neutrino mixing between $\nu_e$ and $\nu_\mu$,
$$\sin\theta_{e\mu} \simeq ({1}/{2})\sqrt{{m_e}/{m_\mu}}
\simeq 0.035 \ ,$$
which is in good agreement with the value $\sin\theta_{e\mu}\simeq 0.04$
($\sin^2 2\theta_{e\mu}\simeq 7\times 10^{-3}$) suggested by
the GALLEX data [13].
However, in the present stage, we do not have any unified understanding of
$b_f$ and $\beta_f$, i.e., they are nothing more than
phenomenological parameters.


Finally, we comment on physical Higgs bosons $\phi_L$ in the present scenario.
We define three mixing states among $\phi_1^1$,
$\phi_2^2$ and $\phi_3^3$ as follows:
$$ \phi'_3=({1}/{\overline{v}})\left(v_1\phi_1^1+v_2\phi_2^2+v_3\phi_3^3
\right)\ ,$$
$$\phi'_1=({1}/{\overline{v}})\sqrt{{2}/{3}}
\left[ (v_3-v_2)\phi_1^1+(v_1-v_3)\phi_2^2+(v_2-v_1)\phi_3^3\right]\ , $$
and $\phi'_2$ are a state which is orthogonal to $\phi'_1$ and $\phi'_3$,
where $\langle\phi^0\rangle={\rm diag}(v_1,v_2,v_3)$ and
$\overline{v}=(v_1^2+v_2^2+v_3^2)^{1/2}$.
Moreover, for convenience, we rewrite $\phi^+$ and $\phi^0$ as
$$\phi^+=i\Pi^+\ ,\ \ \ \phi^0=(H^0-i\Pi^0)/{\sqrt{2}}\ , $$
Then, for the $\Pi^\pm$ and $\Pi^0$ states, the spontaneous symmetry
breaking $\langle\phi^0\rangle \neq 0$ provides nonvanishing masses
except for $\Pi_3^{\prime\pm}$ and $\Pi_3^{\prime 0}$,
which are eaten by gauge bosons $W^\pm$ and $Z^0$.
For $H^0$ states, the potential (4) provides nonvanishing masses
only for the states $H_2^{\prime 0}$ and $H_3^{\prime 0}$.
We do not consider any gauge bosons which eat the massless scalar fields
$H_1^{\prime 0}$ and $H^{j0}_i$ ($i\neq j$), so that
the scalar bosons can  appear as physical
massless Higgs bosons (hereafter, we denote them as $h^0$).
Since the massless bosons $h^0$ cannot couple with $\overline{f}f$,
$A_\mu A^\mu$, $W_\mu^-W^{+\mu}$ and $Z_\mu Z^\mu$ at tree level,
they are harmless to phenomenology in the present accelerator experiments.

The massive Higgs boson $H_3^{\prime 0}$ corresponds to the physical
neutral Higgs boson in the standard model, and it can couple with
$W_\mu^-W^{+\mu}$ and $Z_\mu Z^\mu$.
We suppose that massive physical bosons $\Pi^\pm$, $\Pi^0$,
$H_2^{\prime 0}$ and $H_3^{\prime 0}$ are heavier than weak bosons $W^\pm$
and $Z^0$.
Therefore, a typical production mode of our physical Higgs bosons
is $e^++e^-\rightarrow Z^*\rightarrow H_3^{\prime 0}+Z^*\rightarrow
H_3^{\prime 0}+f+\overline{f}$, where $f\overline{f}$ denotes
$b\overline{b}$, $\tau\overline{\tau}$, $\mu\overline{\mu}$, and so on,
and $Z^*$ means a virtual $Z^0$.
Since we suppose that heavy fermions $F$ are extremely heavy in comparison
with weak bosons and massive physical Higgs bosons $\phi_L$,
our physical Higgs bosons can decay neither into $f\overline{f}$
(at tree level) nor into $f\overline{F}$ ($F\overline{f}$),
so that the dominant decay modes of $H_3^{\prime 0}$ are
$H_3^{\prime 0}\rightarrow h^0\overline{h}^0$ ($h^0=H_1^{\prime 0},
H^{j0}_i$).
As a result, we will observe a characteristic production mode
of $H_3^{\prime 0}$;
$$e^++e^-\rightarrow Z^*\rightarrow Z^* + H_3^{\prime 0} \rightarrow
f\overline{f}+({\rm neutral \ particles})\ . $$
For more phenomenology of the physical Higgs bosons
$\Pi^\pm$, $\Pi^0$ and $H^0$, we will discuss elsewhere.


In conclusion,
we have proposed a unified mass matrix model of quarks and
leptons with seesaw-type matrix form (2)
on the basis of a Higgs mechanism scenario for U(3)-family nonet bosons.
The Higgs potential (4) can lead to the charged lepton mass relation (1)
when we suppose $O_e={\bf 1}$, which is provided by the U(3)-family
singlet Higgs boson $\Phi_0$.
On the other hand, the matrix form $O_q$ in quark mass matrix $M_q$, (8),
has been chosen from a phenomenological consideration.
(We will need further plausible explanation on the reason why there is
such the Higgs boson $\Phi_X$ which couples only with S$_3$ symmetric
states of $F_i$.)
Then, quark mass ratios and KM matrix elements can be fitted only by three
parameters $b_u$, $b_d$ and $\beta_d$ fairly well.
It is worth while that we can obtain a large ratio of $m_t/m_b$
together with a reasonable ratio $m_u/m_d$ without taking  so
hierarchically different values of $b_u$ and $b_d$, i.e.,
with taking $b_d/b_u\simeq 3$.
It should also be noted that since our SU(2)$_L$ doublet Higgs bosons
$\phi_L$ couple only between the ordinary fermions $f_L$
and the heavy fermions $F_R$, the physical Higgs bosons $\phi_L$ cannot
decay into two ordinary fermions at tree level.
If our model is true, it will require re-investigation in experimental
search for physical Higgs bosons.


\vglue.3in

\centerline{\bf Acknowledgments}
The authors are grateful to Professor M.~Tanimoto for reading the
manuscript and stimulating comments.
The authors  also would like to thank Professor T.~Teshima and Professor
R.~Najima for their valuable comments.
This work was supported by the Grant-in-Aid for Scientific Research,
Ministry of Education, Science and Culture, Japan (No.06640407).

\vglue.3in
\newcounter{0000}
\centerline{\bf References and Footnotes}
\begin{list}
{[~\arabic{0000}~]}{\usecounter{0000}
\labelwidth=0.8cm\labelsep=.1cm\setlength{\leftmargin=0.7cm}
{\rightmargin=.2cm}}
\item  ARGUS collaboration, H.~Albrecht {\it et al}.,
Phys.~Lett. {\bf B292}, 221 (1992);
BES collaboration, J.~Z.~Bai {\it et al}., Phys.~Rev.~Lett. {\bf 69},
3021 (1992);
CLEO collaboration, R.~Balest {\it et al}., Phys.~Rev. {\bf D47},
R3671 (1993).
\item Y.~Koide, Lett.~Nuovo Cimento {\bf 34}, 201 (1982); Phys.~Lett.
{\bf B120}, 161 (1983).
\item Y.~Koide, Phys.~Rev. {\bf D28}, 252 (1983).
\item Y.~Koide, Phys.~Rev. {\bf D49}, 2638 (1994).
\item Y.~Koide, Mod.~Phys.~Lett. {\bf A5}, 2319 (1990).
\item M.~Kobayashi and T.~Maskawa, Prog.~Theor.~Phys. {\bf 49}, 652 (1973).
\item
For instance, H.~Harari, H.~Haut and J.~Weyers, Phys.~Lett.
{\bf 78B}, 459 (1978);
H.~Fritzsch and J.~Plankl, Phys.~Lett. {\bf B237}, 451 (1990).
\item C.~Jarlskog, Phys.~Rev.~Lett. {\bf 55}, 1839 (1985);
O.~W.~Greenberg, Phys.~Rev. {\bf D32}, 1841 (1985);
I.~Dunietz, O.~W.~Greenberg, and D.-d.~Wu, Phys.~Rev.~Lett. {\bf 55},
2935 (1985);
C.~Hamzaoui and A.~Barroso, Phys.~Lett. {\bf 154B}, 202 (1985);
D.-d.~Wu, Phys.~Rev. {\bf D33}, 860 (1986).
\item Particle Data Group, K.~Hikasa et al., Phys. Rev. {\bf D45}, S1 (1992).
\item For light quark masses, we have used the values at $\mu=1$ GeV,
$m_u=5.6\pm 1.1$ MeV, $m_d=9.9\pm 1.1$ MeV and $m_s=199\pm 33$ MeV:
C.~A.Dominquez and E.~de Rafael, Anals of Physics {\bf 174}, 372 (1987).
For $m_c$ and $m_b$, we have used the value $m_c(p^2=-m_c^2)=1.26\pm 0.02$
GeV by Narison, and the value $m_b(p^2=m_b^2)=4.72\pm 0.05$ GeV
by Domingnez--Pavaver: S.~Narison, Phys.~Lett. {\bf B216}, 191 (1989);
C.~A.~Dominguez and N.~Paver, Phys.~Lett. {\bf B293}, 197 (1992).
For top quark mass, we have used $m_t(m_t)=174\pm 16$ GeV from CDF
experiment: CDF Collaboration, F.~Abe et al., Preprint
FERIMILAB-PUB-94/097-E (1994).
%
\item We do not consider a grand unification scenario.
We tacitly suppose a composite picture for Higgs bosons and fermions.
The scale $\Lambda_X$ means a scale at which  the Higgs bosons $\Phi_0$
and $\Phi_X$ have their large VEV's, and the U(3)-family singlet boson
$\Phi_0$ couples with all heavy fermions $F$ universally.
\item Y.~Koide, Mod.~Phys.~Lett. {\bf 8}, 2071 (1993).
\item GALLEX collaboration, P.~Anselman {\it et al}., Phys.~Lett. {\bf B285},
390 (1992).
\end{list}

\end{document}